\documentclass[aps,pra,a4paper,twocolumn,showpacs,showkeys]{revtex4}
\usepackage{graphicx,amsmath,bbm,mathrsfs,amssymb,pstricks,times,psfrag}
\newcommand{\ket}[1]{\vert #1 \rangle} \newcommand{\bra}[1]{\langle #1 \vert}
 \newcommand{\sfy}{{\sf y}}

\newcommand{\bmsigma}{\boldsymbol \sigma} \newcommand{\bmSigma}{\boldsymbol \Sigma}
\newcommand{\bmX}{\boldsymbol X} \newcommand{\bmA}{\boldsymbol A}
\newcommand{\bmB}{\boldsymbol B} \newcommand{\bmC}{\boldsymbol C}
\newcommand{\bmS}{\boldsymbol S} \newcommand{\bmLambda}{{\boldsymbol \Lambda}}
 \newcommand{\sfx}{{\sf x}}
\newcommand{\Aop}{{\mathscr A}} 
 \newcommand{\Cop}{{\mathscr C}}
\newcommand{\Xop}{{\mathscr X}} 
 \newcommand{\Top}{{\mathscr T}}

\begin{document}
\title{Selective cloning of Gaussian states by linear optics}
\author{Stefano Olivares} \email{Stefano.Olivares@mi.infn.it}
\affiliation{Dipartimento di Fisica dell'Universit\`a degli Studi di Milano,
Italia.}
\begin{abstract}
We investigate the performances of a selective cloning machine based on
linear optical elements and Gaussian measurements, which allows to clone
at will one of the two incoming input states. This machine is a
complete generalization of a $1\to 2$ cloning scheme demonstrated by
U.~L.~Andersen {\em et al.} [Phys. Rev. Lett. {\bf 94}, 240503 (2005)].
The input-output fidelity is studied for generic Gaussian input state and
the effect of non-unit quantum efficiency is also taken into account.
We show that if the states to be cloned are squeezed states with known
squeezing parameter, then the fidelity can be enhanced using a third suitable
squeezed state during the final stage of the cloning process. A binary
communication protocol based on the selective cloning machne is also discussed.
\end{abstract}
\date{\today}
\maketitle
\section{Introduction}\label{s:intro}
Basic laws of quantum mechanics do not allow the generation of exactly
alike copies of an unknown quantum state
\cite{wooters82.nat,dieks82.pla,cl3,cl4}.  However, approximate copies
can be obtained by using devices called quantum cloning
machine~\cite{buzek96.pra}. The first of such devices was studied to
deal with qubits and then a continuous variable
(CV)~\cite{braunstein05.rev} analog was
developed~\cite{cl:cerf,cerf:PRA:2000}. Thereafter, CV optimal
Gaussian cloners of coherent states based on two quite different
approaches were proposed: the one relies on a single phase insensitive
parametric amplifier \cite{braunstein01.prl,fiurasek01.prl}, the
other, which has been also experimentally realized, is built around a
feed-forward loop \cite{andersen05.prl}. On the other hand, the latter
is much simpler than the first one, overcoming the difficulty of
implementing an efficient phase insensitive amplifier operating at the
fundamental limit. Since the setup of this device is based only on linear
components, throughout this paper we'll refer to it as {\em linear cloning
machine}. Ref.~\cite{OPA:PRA:06} investigated the
performances of the linear cloning machine when the input state was a
single generic Gaussian state (coherent, squeezed coherent or
displaced thermal state) taking into account the effect of fluctuation
of the input state covariance matrix, variation in the setups beam
splitter ratios and losses in the detection scheme.
\par
The aim of this paper is to show that the protocol used by the
linear cloning machine to clone a single input Gaussian state can be
generalized in order achieve the {\em selective} cloning a state
chosen between two inputs.  The possibility to select one of two
states may have useful implementation in binary communication systems
where the two bits are encoded in two quantum states and the goal of
the communication is to sent the information from one sender to two
receivers. We'll address this problem in the final part of the paper.
\par
The paper is structured as follows: in Sec.~\ref{s:scheme} we describe the
selective cloning machine and describe the evolution of the input states
by means of the characteristic function approach. In Sec.~\ref{s:sel}
the requirements of selective symmetric cloning are exploited and
the input-output fidelity is studied. Sec.~\ref{s:enhanced} investigate
the possibility to enhance the cloning fidelity and in Sec.~\ref{s:bin}
a possible application of the selective cloning machine to $1\to 2$
binary communication is proposed. Finally, Sec.~\ref{s:CR} closes the
paper with some concluding remarks.

\section{The selective linear cloning machine}\label{s:scheme}
\begin{figure}[tb]
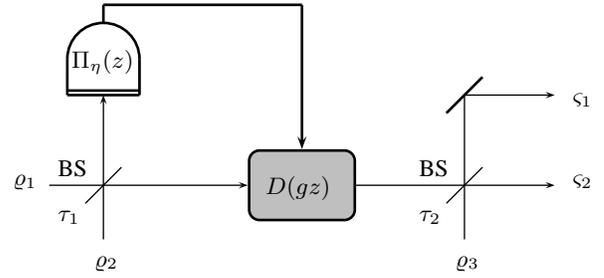

\begin{center}
\psset{unit=1.2cm}
\pspicture(0,0)(6.5,3.2)
\psline[linewidth=0.5pt]{->}(0.4,1)(2.6,1)
\psline[linewidth=0.5pt]{->}(1,0.4)(1,2)
\psline[linewidth=0.5pt]{->}(3.8,1)(6,1)
\psline[linewidth=0.5pt]{->}(5,2)(6,2)
\psline[linewidth=0.5pt]{-}(5,0.4)(5,2)
\psframe[framearc=0.3,linewidth=1pt,fillstyle=solid,fillcolor=lightgray](2.6,0.6)(3.8,1.4)
\put(2.8,0.9){$D(gz)$}
\psline[linewidth=1pt]{-}(1,2.8)(1,3)
\psline[linewidth=1pt]{-}(1,3)(3.2,3)
\psline[linewidth=1pt]{->}(3.2,3)(3.2,1.4)
\psarc[linewidth=1pt](1,2.4){0.4}{0}{180}
\psline[linewidth=1pt]{-}(0.6,2.4)(0.6,2)
\psline[linewidth=1pt]{-}(1.4,2.4)(1.4,2)
\psline[linewidth=1pt]{-}(0.6,2)(1.4,2)
\psline[linewidth=1pt]{-}(0.6,2.05)(1.4,2.05)
\put(0.65,2.3){$\Pi_\eta(z)$}
\psline[linewidth=0.5pt]{-}(0.8,0.8)(1.2,1.2)
\put(0.5,1.1){\small BS}
\put(0.5,0.6){$\tau_1$}
\psline[linewidth=0.5pt]{-}(4.8,0.8)(5.2,1.2)
\put(4.5,1.1){\small BS}
\put(4.5,0.6){$\tau_2$}
\psline[linewidth=1pt]{-}(4.8,1.8)(5.2,2.2)
\put(0,1){$\varrho_{1}$}
\put(0.9,0.1){$\varrho_{2}$}
\put(4.9,0.1){$\varrho_{3}$}
\put(6.2,1.9){$\varsigma_{1}$}
\put(6.2,1){$\varsigma_{2}$}
\endpspicture
\end{center}
\vspace{-.6cm}
\caption{\label{f:cl:scheme} Selective cloning of Gaussian
states by linear optics: the two input states $\varrho_{k}$, $k=1,2$ are
mixed at a beam splitter (BS) of transmissivity $\tau_{1}$. One of the
two emerging beams is measured by a measurement described by the POVM
$\Pi_\eta(z)$ and the outcome $z$ is forwarded to a modulator, which
imposes a  displacement $g z$ on the other outgoing beam, $g$ being a
suitable amplification factor. Finally, the displaced state is  mixed
with the state $\varrho_{3}$ at second beam splitter of transmissivity
$\tau_{2}$. The two outputs, $\varsigma_1$ and $\varsigma_2$,  from
the beam splitter represents the two clones, which may be made
approximately equal to either $\varrho_1$ or $\varrho_2$ by changing
the gain $g$ from $+1$ to $-1$.}
\end{figure}
The selective cloning machine based on linear optics and Gaussian
measurement is schematically depicted in Fig.~\ref{f:cl:scheme}. 
Two input states, denoted by the density operators $\varrho_{k}$, $k=1,2$,
are mixed at a beam splitter (BS) with transmissivity $\tau_1$.
A Gaussian measurement with quantum efficiency $\eta$ is performed on
one of the outgoing beams, the outcome of the measurement being the complex
number $z$. According to these outcomes, the other beam undergoes a
displacement by an amount $gz$, where $g$ is a suitable electronic amplification factor,
and, finally, the two output states, denoted by the density operators $\varsigma_1$
and $\varsigma_2$, are obtained by dividing the displaced state using another
BS with transmissivity $\tau_2$.  When $\tau_1 = \tau_2 = 1/2$,
$g = 1$, $\eta = 1$, $\varrho_2 = \varrho_3 = \ket{0}\bra{0}$
and the Gaussian measurement is an ideal double homodyne detection
the scheme reduces to that of Ref.~\cite{andersen05.prl}, which was
shown to be optimal for Gaussian cloning of coherent states and has been
investigated in Ref.~\cite{OPA:PRA:06,OPA:APU:06,OP:JPA:07}.
In the following we carry out a thorough description of the selective
cloning machine using the characteristic function approach.  
\par
The characteristic function $\chi_{k}(\bmLambda_k) \equiv
\chi[\varrho_{k}](\bmLambda_k)$ associated with a Gaussian
state $\varrho_{k}$ of mode $k=1,2,3$ (see Fig.~\ref{f:cl:scheme}) reads:
\begin{equation}\label{rho:in}
\chi_{k}(\bmLambda_k) = \exp\left\{ -\mbox{$\frac12$} \bmLambda_k^T
\bmsigma_{k}\,\bmLambda_k - i \bmLambda_k^T \bmX_{k}\right\}\,,
\end{equation}
where $\bmLambda_k = (\sfx_k, \sfy_k)^T$, $(\cdots)^T$ denotes the
transposition operation, $\bmsigma_k$ is the covariance matrix, and
$\bmX_{k} = {\rm Tr}[\varrho_{k}\, (\hat x, \hat y)^T]$ is the vector
of mean values, $\hat x$ and $\hat y$ being the quadrature operators
$\hat x = \frac{1}{\sqrt{2}}(\hat{a}+\hat{a}^{\dag})$ and $\hat y =
\frac{1}{i\sqrt{2}}(\hat{a} - \hat{a}^{\dag})$, with $\hat{a}$ and
$\hat{a}^\dagger$ being the field annihilation and creation operator.
In turn, the initial two-mode state $\varrho = \varrho_{1} \otimes
\varrho_{2}$ is Gaussian and its two-mode characteristic function reads:
\begin{equation}
\chi[\varrho](\bmLambda)
= \exp\left\{ -\mbox{$\frac12$} \bmLambda^T
\bmsigma \,\bmLambda - i \bmLambda^T \bmX \right\}\,,
\end{equation}
with
\begin{equation}
\bmsigma = \left(
\begin{array}{c|c}
\bmsigma_{1} & {\boldsymbol 0} \\
\hline
{\boldsymbol 0} & \bmsigma_{2}
\end{array}
\right)\,,\qquad
\bmX = (\bmX_{1}, \bmX_{2})^T\,,
\end{equation}
and $\bmLambda = (\bmLambda_1, \bmLambda_2)$.
Under the action of the first BS the state $\chi[\varrho](\bmLambda)$
preserves its Gaussian form, namely
\begin{equation}
\chi[\varrho](\bmLambda) \rightsquigarrow \chi[\varrho'](\bmLambda) 
= \exp\left\{ -\mbox{$\frac12$} \bmLambda^T
\bmsigma \,\bmLambda - i \bmLambda^T \bmX \right\}\,,
\end{equation}
where $\varrho' = U_{{\rm BS},1}\,\varrho_{1}\otimes\varrho_{2}\,U_{{\rm
BS},1}^{\dag}$, while its covariance matrix and mean values 
transform as~\cite{FOP:napoli:05}:
\begin{align}
&\bmsigma \rightsquigarrow \tilde{\bmsigma} \equiv
{\bmS}_{{\rm BS},1}^T\, \bmsigma\,{\bmS}_{{\rm BS},1} =
\left(
\begin{array}{c|c}
\bmA & \bmC\\
\hline
\bmC^T & \bmB
\end{array}
\right)\,,\label{transf:cvm}\\
&\bmX \rightsquigarrow
\tilde{\bmX} \equiv {\bmS}_{{\rm BS},1}^T\,\bmX =
(\tilde{\bmX}_1,\tilde{\bmX}_2)^T\,,
\label{transf:ave}
\end{align}
$\bmA$, $\bmB$, and $\bmC$ are $2 \times 2$ matrices, and
\begin{equation}\label{symp:BS}
\bmS_{{\rm BS},1} =
\left(
\begin{array}{c|c}
\sqrt{\tau_1}\, \mathbbm{1}_2 & \sqrt{1-\tau_1}\, \mathbbm{1}_2 \\
\hline
-\sqrt{1-\tau_1}\, \mathbbm{1}_2  & \sqrt{\tau_1}\, \mathbbm{1}_2
\end{array}
\right)\,,
\end{equation}
is the symplectic transformation associated with the evolution operator
$U_{{\rm BS},1}$ of the BS with transmission $\tau_1$.  
Note that $\varrho'$ is an entangled state if the set of 
states to be cloned consists of non-classical states, i.e., 
states with singular Glauber P-function or negative Wigner 
function \cite{visent,wang}.
\par
The Gaussian measurement with quantum efficiency $\eta$
(see Fig.~\ref{f:cl:scheme}) is described by the
characteristic function
\begin{equation}
\chi[\Pi_{\eta}(z)](\bmLambda_2) =\frac{1}{\pi}
\exp\left\{ -\mbox{$\frac12$} \bmLambda_2^T\,\bmsigma_{\rm M}\,\bmLambda_2 -
i \bmLambda_2^T\,\bmX_{\rm M} \right\}\,,
\end{equation}
with $\bmX_{\rm M} =\sqrt{2} \left({\rm Re}[z],{\rm Im}[z]\right)^T$
and $\bmsigma_{\rm M} \equiv \bmsigma_{\rm M}(\eta)$.
The probability of obtaining the outcome $z$ is then given by
\begin{align}
p_{\eta}(z) &= {\rm Tr}_{12}[\varrho'\,
\mathbb{I}\otimes\Pi_{\eta}(z)]\label{p:DHD}\\
&= \frac{1}{(2\pi)^2} \int_{\mathbb{R}^4}\!\!\! d^4\bmLambda\,
\chi[\varrho'](\bmLambda)\,
\chi[\mathbb{I}\otimes\Pi_{\eta}(z)](-\bmLambda)\\
&=\frac{
\exp\left\{ -\mbox{$\frac12$}(\bmX_{\rm M}-\tilde{\bmX}_{2})^T\,
\bmSigma^{-1}\,(\bmX_{\rm M}-\tilde{\bmX}_{2})\right\}}
{\pi \sqrt{{\rm Det}[\bmSigma]}}\,,
\end{align}
where $\chi[\mathbb{I}\otimes\Pi_{\eta}(z)](\bmLambda)\equiv
\chi[\mathbb{I}](\bmLambda_1)\,
\chi[\Pi_{\eta}(z)](\bmLambda_2)$, 
$\chi[\mathbb{I}](\bmLambda_1) =
2\pi \delta^{(2)}(\bmLambda_1)$ and $\delta^{(2)}(\zeta)$ is the complex
Dirac's delta function. We also introduced the $2\times 2$ matrix
$\bmSigma = \bmB + \bmsigma_{\rm M}$. 
\par
The conditional state $\varrho_{\rm c}$ of the other outgoing beam, obtained
when the outcome of the measurement is $z$, i.e.,
\begin{equation}
\varrho_{\rm c} =
\frac{{\rm Tr}_{2}[\varrho'\,\Pi_{\eta}(z)]}{p_{\eta}(z)}\,,
\end{equation}
has the following characteristic function (for the sake of clarity we
explicitly write the dependence on $\bmLambda_1$ and $\bmLambda_2$)
\begin{align}
\chi[\varrho_{\rm c}](\bmLambda_1)
=&
\int_{\mathbb{R}^2}\!\!\! d^2\bmLambda_2\,
\frac{
\chi[\varrho'](\bmLambda_1,\bmLambda_2)\,
\chi[\Pi_{\eta}(z)](-\bmLambda_2)}
{p_{\eta}(z)}\\
=&\exp\left\{
-\mbox{$\frac12$}\bmLambda_1^T
\left[ \bmA - \bmC\bmSigma^{-1}\bmC^T \right]
\bmLambda_1 \right.\nonumber\\
&\left.
-i \bmLambda_1^T \left[ \bmC\bmSigma^{-1}
\left(\bmX_{\rm M}-\tilde{\bmX}_{2}\right) + \tilde{\bmX}_1 \right]\right\}\,.
\end{align}
Now, the conditional state $\varrho_{\rm c}$ is displaced by the amount
$gz$ resulting from the measurement amplified by a factor $g$.
By averaging over all possible outcomes of the double-homodyne
detection, we obtain the following output state:
\begin{equation}\label{rho:d}
\varrho_{\rm d} = \int_{\mathbb{C}}d^2 z\,
p_{\eta}(z)\,D(gz)\, \varrho_{c} \,D^{\dag}(gz)\,,
\end{equation}
with $D(\zeta)$ being the displacement operator. In turn, the characteristic
function reads as follows:
\begin{equation}
\chi[\varrho_{\rm d}](\bmLambda_1) = \exp\left\{
-\mbox{$\frac12$} \bmLambda_1^T\,\bmsigma_{\rm d}\,\bmLambda_1
-i \bmLambda_1^T \bmX_{\rm d} \right\}\,,
\end{equation}
with $\bmsigma_{\rm d} = \bmA + g^2\bmSigma + g(\bmC+\bmC^T)$
and $\bmX_{\rm d} = \tilde{\bmX}_1 + g \tilde{\bmX}_2$.
The conditional state (\ref{rho:d}) is then sent to a second BS
with transmission $\tau_2$ (see Fig.~\ref{f:cl:scheme}),
where it is mixed with the Gaussian state $\varrho_{3}$, and finally
the two clones are generated. Note that, in practice, the average
over all the possible outcomes $z$ in Eq.~(\ref{rho:d})
should be performed at this stage, that is after the second BS.
On the other hand, because of the linearity of the
integration, the results are identical, but performing the
averaging just before the BS simplifies the
calculations. Since $\varrho_{\rm d}$ is still Gaussian, the
two-mode state $\varrho_{\rm f} = \varrho_{\rm d} \otimes
\varrho_3$ is a Gaussian with covariance matrix and mean given by
\begin{equation}
\bmsigma_{\rm f} = \left(
\begin{array}{c|c}
\bmsigma_{\rm d} & {\boldsymbol 0} \\
\hline
{\boldsymbol 0} & \bmsigma_3
\end{array}
\right)\,,\qquad
\bmX_{\rm f} = (\bmX_{\rm d}, \bmX_3)^T\,,
\end{equation}
respectively, which, as in the case of Eqs.~(\ref{transf:cvm}) and
(\ref{transf:ave}), under the action of the BS transform as follows:
\begin{align}
&\bmsigma_{\rm f} \rightsquigarrow \bmsigma_{\rm out} \equiv
{\bmS}_{{\rm BS},2}^T\, \bmsigma_{\rm f}\,{\bmS}_{{\rm BS},2} =
\left(
\begin{array}{c|c}
\Aop_1 & \Cop\\
\hline
\Cop^T & \Aop_2
\end{array}
\right)\,,\label{transf:cvm:fin}\\
&\bmX_{\rm f} \rightsquigarrow
\bmX_{\rm out} \equiv {\bmS}_{{\rm BS},2}^T\,\bmX_{\rm f} =
(\Xop_1,\Xop_2)^T\,,
\label{transf:ave:fin}
\end{align}
where $\Aop_k$ and $\Cop$ are $2 \times 2$ matrices, and $\bmS_{{\rm BS},2}$ is
the symplectic matrix given by Eq.~(\ref{symp:BS}) with $\tau_1$ replaced by 
$\tau_2$. Finally, the (Gaussian) characteristic function of the
clone $\varsigma_k$, $k=1,2$, is obtained by integrating over $\bmLambda_{h}$,
$h\ne k$, the two-mode characteristic function $\chi[\varrho_{\rm
out}](\bmLambda_1,\bmLambda_2)$, where $\varrho_{\rm out} = U_{{\rm BS},2}\,
\varrho_{\rm f}\otimes \varrho_3 \,U_{{\rm BS},2}^{\dag}$, i.e.,
\begin{align}
\chi[\varsigma_k](\bmLambda_k) &= \frac{1}{2\pi}
\int_{\mathbb{R}^2}\!\!\! d^2\bmLambda_h\,
\chi[\varrho_{\rm out}](\bmLambda_1,\bmLambda_2)\\
&=\exp\left\{ -\mbox{$\frac12$}\bmLambda_{k}^T\,\Aop_k\,\bmLambda_{k}
- i \bmLambda_k^T\, \Xop_k\right\}\,.\label{clone:k}
\end{align}
The explicit expressions of $\Xop_1$ and $\Xop_2$ are
\begin{subequations}\label{X:k}
\begin{align}
\Xop_1 =& \sqrt{\tau_2}\left( f_1 \bmX_{1} +
 f_2 \bmX_{2} \right) -\sqrt{1-\tau_2} \bmX_3
\,,\\
\Xop_2 =& \sqrt{1-\tau_2}\left( f_1 \bmX_{1} +
 f_2 \bmX_{2} \right) + \sqrt{\tau_2} \bmX_3 \,.
\end{align}
\end{subequations}
with
\begin{align}
f_1 \equiv f_1(\tau_1,\tau_2,g) =& \sqrt{\tau_1} + g\sqrt{1-\tau_1}\,,\\
f_2 \equiv f_2(\tau_1,\tau_2,g) =& g\sqrt{\tau_1} -\sqrt{1-\tau_1}\,,
\end{align}
whereas, $\Aop_{1}$ and $\Aop_{2}$ can be written in a compact form as
follows:
\begin{subequations}\label{A:k}
\begin{align}\Aop_{1} =& \tau_2 \left( f_1^2 \bmsigma_{1} +
 f_2^2 \bmsigma_{2} + g^2\, \bmsigma_{\rm M} \right) 
+ (1-\tau_2)\,\bmsigma_{3}\,,\\
\Aop_{2} =& (1-\tau_2) \left( f_1^2 \bmsigma_{1} +
f_2^2 \bmsigma_{2} + g^2\, \bmsigma_{\rm M} \right)
+ \tau_2\,\bmsigma_{3} \,.
\end{align}
\end{subequations}

\section{Selective Cloning}\label{s:sel}
From Eqs.~(\ref{X:k}) and (\ref{A:k}) we see that the two
outgoing states $\varsigma_1$ and $\varsigma_2$ are generally different.
In this paper we'll consider the case in which the clones are equal, therefore,
In order to make them exactly alike, one have to put $\tau_2=1/2$ and
$\bmX_3={\boldsymbol 0}$: in this case, $\Xop_1 = \Xop_2$ and $\Aop_1=\Aop_2$.
A further inspection of Eqs.~(\ref{X:k}) and (\ref{A:k}) with $\tau_2 = 1/2$,
shows that the states $\varsigma_k$ could be quite different from both the
input states, being the covariance matrices and the mean values vectors a
linear combination of the input ones. On the other hand, if $f_2$
(or $f_1$) vanishes, then the Gaussian output states depend only on
$\bmsigma_1$, $\bmX_1$ (or $\bmsigma_2$, $\bmX_2$), $\bmsigma_3$ and
$\bmsigma_{\rm M}$. In the following we'll investigate thoroughly this
scenario.
\par
After we have chosen the symmetric outputs setup, i.e., $\tau_2=1/2$
and $\bmX_3={\boldsymbol 0}$, we are interested in removing the dependence
on the state, e.g., $\varrho_2$ from the output states, namely, we want
to let $f_2$ vanish; this is achieved when
\begin{equation}
g\equiv g_1(\tau_1)=\sqrt{(1-\tau_1)/\tau_1}\,,
\end{equation}
which gives $f_1=\tau_1^{-1/2}$ and leads to
\begin{align}
\Xop_1&=\Xop_2=(2\tau_1)^{-1/2}\bmX_1 \\
\Aop_1&=\Aop_2=\frac12 \left[
\frac{1}{\tau_1}\,\bmsigma_1 + \bmsigma_3 +
\frac{1-\tau_1}{\tau_1}\,\bmsigma_{\rm M}
\right]\,.
\end{align}
It is now clear that if the first BS is balanced ($\tau_1=1/2$), we obtain
\begin{subequations}\label{1:cl}
\begin{align}
\Xop_1&=\Xop_2=\bmX_1 \\
\Aop_1&=\Aop_2=\bmsigma_1 + \frac12 \left(
\bmsigma_3 + \bmsigma_{\rm M}
\right)\,.
\end{align}
\end{subequations}
This is the $1\to 2$ symmetric cloning of the state $\varrho_1$. This
configuration has been experimentally implemented to optimally clone
coherent states \cite{andersen05.prl,OPA:PRA:06}. Notice that $g_1(1/2)=1$.
\par
On the contrary, in order to eliminate the dependence on the state
$\varrho_1$ one needs (we are assuming again $\tau_2=1/2$
and $\bmX_3={\boldsymbol 0}$):
\begin{equation}
g\equiv g_2(\tau_1)=-\sqrt{\tau_1/(1-\tau_1)}\,,
\end{equation}
which gives $f_2=-(1-\tau_1)^{-1/2}$ and leads to
\begin{align}
\Xop_1&=\Xop_2=-[2(1-\tau_1)]^{-1/2}\bmX_2 \\
\Aop_1&=\Aop_2=\frac12 \left[
\frac{1}{1-\tau_1}\,\bmsigma_2 + \bmsigma_3 +
\frac{\tau_1}{1-\tau_1}\,\bmsigma_{\rm M}
\right]\,,
\end{align}
and if $\tau_1=1/2$ one has
\begin{subequations}\label{2:cl}
\begin{align}
\Xop_1&=\Xop_2=-\bmX_2 \\
\Aop_1&=\Aop_2=\bmsigma_2 + \frac12 \left(
\bmsigma_3 + \bmsigma_{\rm M}
\right)\,.
\end{align}
\end{subequations}
As a matter of fact, to obtain the actual symmetric cloning of the state
$\varrho_2$ we have to implement an unitary transformation to change the phase
of the output states as follows: $\Xop_h \to -\Xop_h$.  Notice that
$g_2(1/2)=-1$.
\par
The results of this Section are summarized in Table \ref{t:sel}: in the case of
symmetric cloning ($\tau_1=\tau_2 = 1/2$ and $\bmX_3={\boldsymbol 0}$), one can
select the state to clone simply change the value of the gain $g$ from $+1$ to
$-1$.
\begin{table}
\caption{\label{t:sel} Selective symmetric cloning ($\tau_1=\tau_2 = 1/2$ and
$\bmX_3={\boldsymbol 0}$): changing the value of the electronic gain from
$+1$ to $-1$ one can choose to clone the state $\varrho_1$ or $\varrho_2$
respectively. Notice that if $g=-1$ a unitary transformation at the output
is needed in order to obtain the right sign of the amplitude $\Xop_k$, $k=1,2$.}
\begin{ruledtabular}
\begin{tabular}{ccc}
$g$ & $\Aop_1=\Aop_2$ & $\Xop_1 = \Xop_2$\\
\hline
$+1$ & $\bmsigma_1 + \frac12 \left( \bmsigma_3 + \bmsigma_{\rm M} \right) $ & $\bmX_1$\\
$-1$ & $\bmsigma_2 + \frac12 \left( \bmsigma_3 + \bmsigma_{\rm M} \right)$ & $-\bmX_2$
\end{tabular}
\end{ruledtabular}
\end{table}

\section{Enhancement of linear cloning fidelity}\label{s:enhanced}
The similarity between the input state $\varrho_k$ and the clone $\varsigma_h$,
$k,h=1,2$, can be quantified by means of the fidelity \cite{uhlman:RepMP:76}
\begin{equation}
F(\varrho_k,\varsigma_h) =
\left({\rm Tr}\left[
\sqrt{\sqrt{\varrho_{k}}\,\varsigma_{h}\sqrt{\varrho_{k}}}
\right]\right)^2\,,
\end{equation}
which, for Gaussian states, reduces to
 \cite{OPA:PRA:06,scutaru:JPA:98,nha:2005} 
\begin{align}\label{gen:fid}
&{F}_{\eta} \equiv {F}(\varrho_{k},\varsigma_h) =
\frac{1}
{
\sqrt{{\rm Det}[\bmsigma_{k}+\Aop_h]+\delta}-\sqrt{\delta}
}\nonumber\\
&\times\exp\left\{
-\mbox{$\frac12$} (\bmX_{k}-\Xop_h)^T(\bmsigma_{k}+\Aop_{h})^{-1}
(\bmX_{k}-\Xop_h)
\right\}
\,,
\end{align}
where $\delta = 4({\rm Det}[\bmsigma_{k}]-\frac14)
({\rm Det}[\Aop_h]-\frac14)$. Note that for pure Gaussian
states ${\rm Det}[\bmsigma_{k}] = \frac14$, and in turn $\delta = 0$.
In the case of symmetric cloning $\bmX_k=\Xop_h$, the fidelity
(\ref{gen:fid}) reduces to
\begin{align}\label{symm:fid}
F_\eta(\bmsigma_k,\bmsigma_3,\bmsigma_{\rm M}) \equiv \frac{1}
{\sqrt{{\rm Det}[\bmsigma_{k}+\Aop_h]+\delta}-\sqrt{\delta}}\,,
\end{align}
and the cloning machine is said to be {\em universal} because of its invariance
with respect to displacement of the input states.
\par
It is a matter of fact that we can now maximize Eq.~(\ref{symm:fid})
by a suitable choice of the state $\varrho_3$ ($\bmsigma_1$,
$\bmsigma_2$, and $\bmsigma_{\rm M}$ being fixed).
Without loss of generality we assume that the covariance matrix associated
with $\varrho_3$ has the following diagonal form
\begin{equation}
\bmsigma_3 = \left(\begin{array}{cc}
\omega_{11} & 0 \\
0 & \omega_{22} 
\end{array}
\right)
\end{equation}
with
\begin{equation}
\omega_{11}=\frac{2\Top+1}{2}\,e^{2s},\quad
\omega_{22}=\frac{2\Top+1}{2}\,e^{-2s}\,,
\end{equation}
i.e., a squeezed thermal state with $\Top$ mean thermal photons
and squeezed parameter $s$. We recall that $\bmX_3={\boldsymbol 0}$
in order to fulfill the symmetric cloning requirements. Now, if
\begin{equation}
\bmsigma_k = \left(\begin{array}{cc}
\gamma_{11}^{(k)} & \gamma_{12}^{(k)} \\
\gamma_{12}^{(k)} & \gamma_{22}^{(k)} 
\end{array}
\right)\,,\quad
\bmsigma_{\rm M} = \left(\begin{array}{cc}
\Delta^2_{11} & \Delta_{12} \\
\Delta_{12} & \Delta^{2}_{22} 
\end{array}
\right)\,,
\end{equation}
are the explicit forms of the covariance matrices of $\varrho_{k}$, $k=1,2$,
and of the measurement $\Pi_{\eta}(z)$, respectively, then we find that
the fidelity reaches the maximum for (for the sake of simplicity we do not
report explicitly the dependence of $\gamma_{mn}^{(k)}$ on $k$, being clear
what is the input state $\varrho_{k}$ under consideration)
\begin{equation}\label{opt:cond}
s = \overline{s} \equiv \frac{1}{4} \log\left(
\frac{4\gamma_{11} + \Delta^2_{11}}{4\gamma_{22} + \Delta^2_{22}}
\right)\,,\quad \Top = 0\,,
\end{equation}
i.e., $\varrho_3$ should be a squeezed vacuum state with covariance matrix
$\bmsigma_3\equiv\bmsigma_{\overline{s}} =
\frac12 {\rm Diag}(e^{2\overline{s}},e^{-2\overline{s}})$. Indeed, such a
maximization of the fidelity requires the knowledge of $\gamma_{11}$
and $\gamma_{22}$.
\par
\begin{figure}[tb]
\psfrag{xl}{$r$}
\psfrag{ylx}{$F_\eta$}
\includegraphics{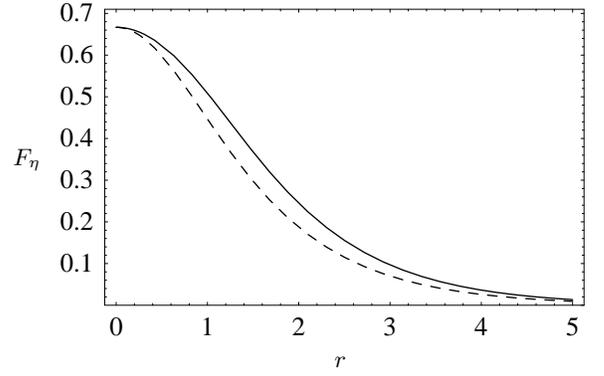}
\vspace{-0.3cm}
\caption{\label{f:Enh:SqB} Plot of the fidelity
$F_\eta(\bmsigma_k,\bmsigma_{3},\bmsigma_{\rm M})$ in the case of symmetric cloning
when $\varrho_1$ is a squeezed state with real squeezing parameter $r$;
$\bmsigma_{3}$ is chosen to be the covariance matrix $\bmsigma_{\overline{s}}$
of a vacuum squeezed state (solid line) or $\bmsigma_{0}$ of the vacuum
state (dashed line). See text for details. We set $\eta=1$.}
\end{figure}
\begin{figure}[tb]
\psfrag{xl}{$r$}
\psfrag{ylx}{$G$}
\includegraphics{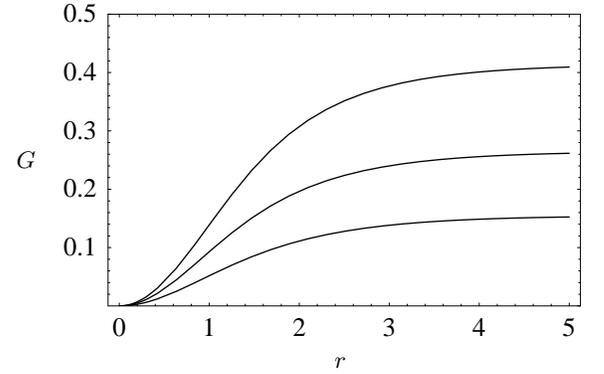}
\vspace{-0.3cm}
\caption{\label{f:Enh:Sq} Plot of the $G(r,\eta)$ given in Eq.~(\ref{sq:enh})
as a function of the squeezing parameter of the input state for different values
of $\eta$, form top to bottom: $\eta=1.0$, $0.75$ and $0.5$. See the text for details.}
\end{figure}

The result obtained above generalizes the conclusions given in Ref.~\cite{OPA:PRA:06}.
The linear cloning machine described in \cite{OPA:PRA:06}, used
to perform $1 \to 2$ cloning of the state $\varrho_1$, follows from the present
scheme choosing $\varrho_2=\varrho_3=\ket{0}\bra{0}$,
corresponding to $\bmsigma_1=\bmsigma_3=\bmsigma_0\equiv\frac12{\mathbbm 1}_2$,
and $\bmsigma_{\rm M}=\frac{2-\eta}{2\eta}{\mathbbm 1}_2$, which describes the
covariance matrix of the double homodyne detection with quantum efficiency
$\eta$. From Eq.~(\ref{opt:cond}) we see that sending the vacuum
into the second BS is the best choice only if $\varrho_1$ is a coherent state or
a displaced thermal state \cite{OPA:PRA:06} (in both the cases $s=0$ and $\bmsigma_3$
reduces to the vacuum state covariance matrix being $\bmX_3={\boldsymbol 0}$).
On the contrary, when $\bmsigma_k$ is the covariance matrix associated with
the squeezed state $D(\alpha)S(r)\ket{0}=\ket{\alpha,r}$, where
$D(\alpha)=\exp\{\alpha a^{\dag} - \alpha^{*}a\}$ and $S(r)=
\exp\{\frac12 r({a^{\dag}}^{2} - a^{2})\}$ are the displacement and squeezing
operators, respectively, $r$ being the real squeezing parameter, then $2\gamma_{11} =
2\gamma_{22}^{-1} = e^{2r}$ and the cloning fidelity is optimized if $\varrho_3$
is a squeezed state with squeezing parameter given by Eq.~(\ref{opt:cond}).
Fig.~\ref{f:Enh:SqB} shows the enhancement of the fidelity in the case
of squeezed state $1\to 2$ cloning when a suitable squeezed vacuum state
with squeezing parameter $\overline{s}$ given in Eq.~(\ref{opt:cond}) is used
instead of the vacuum state as input $\varrho_3$ (see Fig.~\ref{f:cl:scheme}).
The effect of non-unit quantum efficiency can be seen in Fig.~\ref{f:Enh:Sq}
where we plot the quantity
\begin{equation}\label{sq:enh}
G(r,\eta) = \frac{F_\eta(\bmsigma_1,\bmsigma_{\overline{s}},\bmsigma_{\rm M})
-F_\eta(\bmsigma_1,\bmsigma_0,\bmsigma_{\rm M})}
{F_\eta(\bmsigma_1,\bmsigma_0,\bmsigma_{\rm M})}\,,
\end{equation}
as a function of $r$ for different values of $\eta$. $G(r,\eta)$ expresses
the relative improvement of cloning fidelity. As it is apparent from the plot,
one has enhancement of fidelity for any value of $\eta$ as far as the signals
show nonzero squeezing.

\section{$\boldsymbol{1\to 2}$ binary communication}\label{s:bin}
In this Section we address an application of the selective cloning
machine to a $1\to 2$ binary communication protocol. The goal is to
encode a classical sequence (string) $\cal S$ of two classical
symbols, e.g., ``$-1$'' and ``$+1$'', into a quantum sequence $\cal S'$
of two quantum states, e.g., $\varrho_1$ and $\varrho_2$, eventually
unknown, and to send it to two receivers, which are interested not
only in the classical message but also in the quantum states encoding
it. In this case a cloning machine is necessary to generate the copies
${\cal R}_1$ and ${\cal R}_2$ of $\cal S'$. Let us now assume that the
{\em sender}, which possesses the string $\cal S$, is not able to
generate $\cal S'$ himself, so he needs a {\em service provider} that
provides a communication channel based on the states $\varrho_1$ and
$\varrho_2$. However, since the service provider does not know $\cal
S$, the communication channel should be independent on on the message
the sender want to send. In this scenario the selective cloning
machine (operating in the symmetric cloning regime) presented above
can be a useful tool.
\par
\begin{figure}[tb]
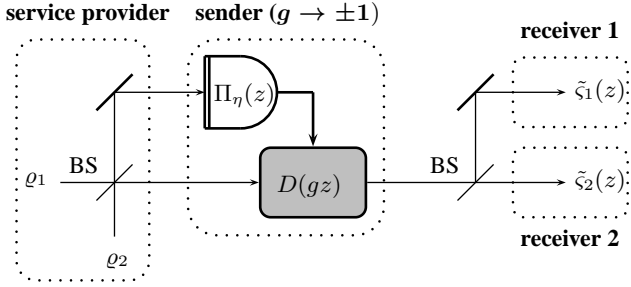

\begin{center}
\psset{unit=1.2cm}
\pspicture(0,-0.1)(6.7,3.2)
%
\psframe[framearc=0.3,linewidth=1pt,linestyle=dotted](-0.1,-0.1)(1.4,2.6)
\put(-0.2,2.8){\bf service provider}
\psframe[framearc=0.3,linewidth=1pt,linestyle=dotted](1.8,0.4)(4,2.6)
\put(1.9,2.8){\bf sender ($\boldsymbol{g\to \pm 1})$}
\psframe[framearc=0.3,linewidth=1pt,linestyle=dotted](5.4,0.6)(6.7,1.4)
\put(5.5,2.6){\bf receiver 1}
\psframe[framearc=0.3,linewidth=1pt,linestyle=dotted](5.4,1.6)(6.7,2.4)
\put(5.5,0.3){\bf receiver 2}
\psline[linewidth=1pt]{-}(0.8,1.8)(1.2,2.2)
\psline[linewidth=0.5pt]{->}(0.4,1)(2.6,1)
\psline[linewidth=0.5pt]{-}(1,0.4)(1,2)
\psline[linewidth=0.5pt]{->}(1,2)(2,2)
\psline[linewidth=0.5pt]{->}(3.8,1)(6,1)
\psline[linewidth=0.5pt]{->}(5,2)(6,2)
\psline[linewidth=0.5pt]{-}(5,1)(5,2)
\psframe[framearc=0.3,linewidth=1pt,fillstyle=solid,fillcolor=lightgray](2.6,0.6)(3.8,1.4)
\put(2.8,0.9){$D(gz)$}
\psline[linewidth=1pt]{-}(2.8,2)(3.2,2)
\psline[linewidth=1pt]{->}(3.2,2)(3.2,1.4)
\psarc[linewidth=1pt](2.4,2.0){0.4}{-90}{90}
\psline[linewidth=1pt]{-}(2.0,1.6)(2.4,1.6)
\psline[linewidth=1pt]{-}(2.0,2.4)(2.4,2.4)
\psline[linewidth=1pt]{-}(2.0,1.6)(2.0,2.4)
\psline[linewidth=1pt]{-}(2.05,1.6)(2.05,2.4)
\put(2.1,1.9){$\Pi_\eta(z)$}
\psline[linewidth=0.5pt]{-}(0.8,0.8)(1.2,1.2)
\put(0.5,1.1){\small BS}
\psline[linewidth=0.5pt]{-}(4.8,0.8)(5.2,1.2)
\put(4.5,1.1){\small BS}
\psline[linewidth=1pt]{-}(4.8,1.8)(5.2,2.2)
\put(0,1){$\varrho_{1}$}
\put(0.9,0.1){$\varrho_{2}$}
\put(6.1,1.95){$\tilde\varsigma_{1}(z)$}
\put(6.1,0.95){$\tilde\varsigma_{2}(z)$}
\endpspicture
\end{center}
\vspace{-.6cm}
\caption{\label{f:binary} $1 \to 2$ binary communication assisted by
the selective cloning machine. The ``service provider'' provides
the communication channel by mixing the two states $\varrho_1$ and
$\varrho_2$ at a balanced BS and by addressing the outgoing beams to the
``sender''. The sender performs a measurement on one of the beams and
displaces the other by an amount $gz$, $z$ being the measurement's outcome
and $g$ being chosen according to the bit the sender wants to encode.
Finally, the message is split into two clones by means of a second BS.
See text for detalis.}
\end{figure}
The $1\to 2$ communication protocol based on the selective cloning machine
is sketched in Fig.~\ref{f:binary} and can be summarized in these steps:
\begin{itemize}
\item the service provider mixes $\varrho_1$ and $\varrho_2$ at the balanced
BS and addresses the outputs to the sender;
\item the sender performs the double homodyne detection onto one of the
two beams and displaces the other one by an amount $g z$, $z$ being the
outcome of the measurement and $g$ being chosen according to the entries
$\pm 1$ of $\cal S$;
\item the displaced beam is divided into the two clones
$\tilde\varsigma_1(z)=\tilde\varsigma_2(z)\equiv\tilde{\varsigma}^{(k)}(z)$
by means of another balanced BS, with $k=1$ ($k=2$) if $g=+1$ ($g=-1$).
\end{itemize}
It is worth noting that the selective cloning machine is now operating
at a ``single shot'' regime, namely, each clone is obtained after a
single outcome $z$ of the double homodyne detection and not after a
complete measurement onto a state. In turn, each clone actually
depends on $z$.  Once the receivers get the single clone, they need a
strategy to decide if the bit was $+1$, corresponding to $\varrho_1$,
or $-1$, corresponding to $\varrho_2$.
\par
\begin{figure}[tb]
\psfrag{xlb}{$\alpha$}
\psfrag{ylb}{$\overline{H}_{\rm e}$}
\includegraphics{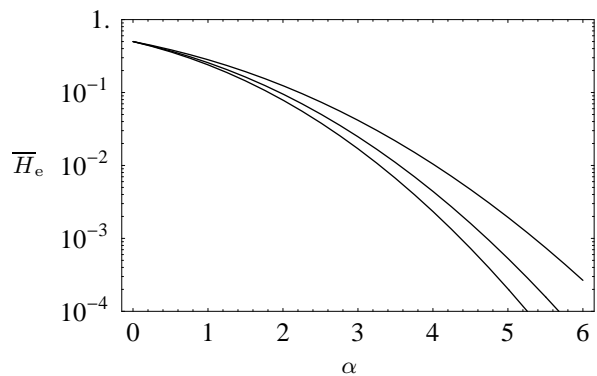}
\vspace{-0.3cm}
\caption{\label{f:comm:eta} Plot of the average error probability
$\overline{H}_{\rm e}(\alpha,\eta,\varepsilon)$ as a function of the
amplitude $\alpha$ and different values of the quantum efficiencies:
we set $\varepsilon = 1.0$ and, from bottom to top, $\eta=1.0$, $0.5$, and $0.75$.}
\end{figure}
\begin{figure}[tb]
\psfrag{xlb}{$\alpha$}
\psfrag{ylb}{$\overline{H}_{\rm e}$}
\includegraphics{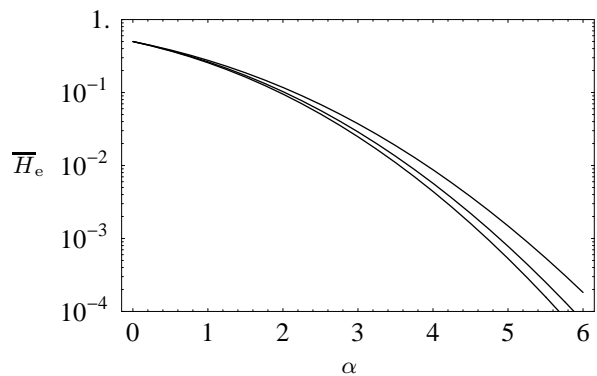}
\vspace{-0.3cm}
\caption{\label{f:comm:epsilon} Plot of the average error probability
$\overline{H}_{\rm e}(\alpha,\eta,\varepsilon)$ as a function of the
amplitude $\alpha$ and different values of the quantum efficiencies:
we set $\eta = 0.75$ and, from bottom to top, $\varepsilon=1.0$, $0.5$, and $0.75$.}
\end{figure}

In order to illustrate the protocol, in the following we address the
simple case in which
\begin{equation}
\varrho_1=\varrho_2=\ket{\alpha}\bra{\alpha}\,,
\end{equation}
are coherent states, i.e., $\bmsigma_k = \frac12 {\mathbbm 1}_2$ and
$\bmX_1=\bmX_2= \sqrt{2}(\alpha,0)$  (for the sake of simplicity we take
$\alpha$ as real and positive). We recall that the clones of $\varrho_2$
have the amplitude with a $\pi$ phase shift (see Table \ref{t:sel}) with
respect to input one: in this way it is possible to distinguish between
$\tilde{\varsigma}^{(1)}(z)$ and $\tilde{\varsigma}^{(2)}(z)$.
Note that one has
\begin{equation}
U_{\rm BS,1}\varrho_1\otimes\varrho_2 U_{\rm BS,1}^{\dag} =
\ket{0}\bra{0}\otimes\ket{\sqrt{2}\alpha}\bra{\sqrt{2}\alpha}\,.
\end{equation}
One of the possible strategies to distinguish between $\tilde\varsigma^{(1)}(z)$
and $\tilde\varsigma^{(2)}(z)$ is performing a homodyne detection, which
is described by the POVM \cite{ComEnt}
\begin{equation}\label{Pi:hom:eta}
\Pi_{x}(\varepsilon) = \frac{1}{\sqrt{2 \pi \sigma_{\varepsilon}^2}}
\int_{\mathbbm R} {\rm d}y \,
\exp\left\{-\frac{(y-x)^2}{2\sigma_{\varepsilon}^2}\right\} \, \Pi_{y}\, ,
\end{equation}
where $\sigma_{\varepsilon}^2 =  (1-\varepsilon)/(4\varepsilon)$,
$\varepsilon$ is the detection
quantum efficiency, and $\Pi_{y}=\ket{y}\bra{y}$, with
\begin{equation}
\ket{y}=\frac{e^{-y^2/2}}{\pi^{1/4}} \sum_{n=0}^{\infty}
\frac{H_n (y)}{\sqrt{n! \, 2^n}}\,\ket{n}\label{xquad}
\end{equation}
being an eigenstate of the quadrature operator $\hat y=\frac{1}{\sqrt{2}}
(a+a^\dag)$ of the
measured mode. In equation (\ref{xquad}) $H_n(y)$ denotes the $n$-th
Hermite polynomials. Finally, the decision is taken according to the following
rule: if $x \ge \overline{x} \Rightarrow k=1$, otherwise $k=2$,
$\overline{x}$ being a threshold value. On the other hand,
$\tilde\varsigma^{(1)}(z)$ and $\tilde\varsigma^{(2)}(z)$ are not orthogonal,
and then we have to evaluate the probability to infer the wrong state,
namely, the {\em error probability} defined as follows:
\begin{equation}\label{err:prob}
H_{\rm e}(z) = \frac12 \left[
P_z(2|1) + P_z(1|2)
\right]\,,
\end{equation}
where $P_z(h|k)$ is the probability to infer the state $\tilde\varsigma^{(h)}(z)$
when the actual state was $\tilde\varsigma^{(k)}(z)$, $h\ne k$. In writing
Eq.~(\ref{err:prob}) we assumed that the two states are sent with the same
{\em a priori} probability $p=1/2$. The explicit expressions of $P_z(2|1)$ and
 $P_z(1|2)$ read as follows:
\begin{subequations}
\begin{align}
P_z(2|1) &= \int_{-\infty}^{\overline{x}}\!\!\!\!\!\!\! dx\, {\rm Tr}\left[
\tilde\varsigma^{(1)}(z)\, \Pi_x(\varepsilon)
\right]\,,\\
P_z(1|2) &= \int^{+\infty}_{\overline{x}}\!\!\!\!\!\!\!\!\! dx\, {\rm Tr}\left[
\tilde\varsigma^{(2)}(z)\, \Pi_x(\varepsilon)
\right]\,.
\end{align}
\end{subequations}
It is easy to see that because of the choice of the states $\varrho_1$ and
$\varrho_2$, the probability $H_{\rm e}(z)$ is minimum when $\overline{x}=0$.
The average error probability is then given by
\begin{equation}\label{ave:error}
\overline{H}_{\rm e}(\alpha,\eta,\varepsilon) =
\int_{\mathbbm C}d^2z\,p_{\eta}(z)\,H_{\rm e}(z)\,,
\end{equation}
where $p_{\eta}(z)$ is the double homodyne detection probability given by
Eq.~(\ref{p:DHD}). In Figs.~\ref{f:comm:eta} and \ref{f:comm:epsilon} we
plot Eq.~(\ref{ave:error}) as a function of the amplitude $\alpha$
and different values of $\eta$ and $\varepsilon$.
As one may expect, in order to reduce the error probability one has to encrease
the amplitude of the input coherent states.
\par
It is worth mentioning that when $\varrho_1$ and $\varrho_2$ are non-classical
state, then $U_{\rm BS,1}\varrho_1\otimes\varrho_2 U_{\rm BS,1}^{\dag}$ is
entangled \cite{visent,wang} and such correlations can be used to
reveal the presence of an eavesdropper along the communication line by means
a suitable non-locality test \cite{bana,nha}.

\section{Concluding remarks}\label{s:CR}
We have addressed the performances of $1\to 2$ selective cloning
machine based on linear optics and Gaussian measurement, which allows
to clone one of two incoming input states. We have shown that
this is achieved simply changing the gain of a feed-forward loop. Moreover
a third Gaussian state can be used in the final stage of the cloning process
in order to enhance the input-output fidelity.
We have found that for coherent or thermal states this state reduces to the
vacuum state, whereas a vacuum squeezed state depending on the squeezing
parameter of the inputs and on the measurement should be considered when the
states to be cloned are squeezed states.
Finally, a protocol for $1\to 2$ binary
communication involving the selective cloning machine has been proposed
and the average error probability has been evaluated for a particular
choice of the involved states.

\section*{Acknowledgments}
Fruitful discussions with M.~G.~A.~Paris and A.~R.~Rossi are kindly acknowledged.
This work has been supported by MIUR through the project PRIN-2005024254-002.
\vfill


\end{document}